\documentstyle[epsf,amsmath]{mn}

\newif\ifAMStwofonts
\ifoldfss
  \ifCUPmtlplainloaded \else
    \NewTextAlphabet{textbfit} {cmbxti10} {}
    \NewTextAlphabet{textbfss} {cmssbx10} {}
    \NewMathAlphabet{mathbfit} {cmbxti10} {} % for math mode
    \NewMathAlphabet{mathbfss} {cmssbx10} {} %  "   "    "
  \fi
  \ifAMStwofonts
    \ifCUPmtlplainloaded \else
      \NewSymbolFont{upmath} {eurm10}
      \NewSymbolFont{AMSa} {msam10}
      \NewMathSymbol{\upi}     {0}{upmath}{19}
      \NewMathSymbol{\umu}     {0}{upmath}{16}
      \NewMathSymbol{\upartial}{0}{upmath}{40}
      \NewMathSymbol{\leqslant}{3}{AMSa}{36}
      \NewMathSymbol{\geqslant}{3}{AMSa}{3E}

       \let\le=\leqslant
      \let\geq=\geqslant 
    \fi
  \fi
\fi % End of OFSS

\ifnfssone
  \newmathalphabet{\mathit}
  \addtoversion{normal}{\mathit}{cmr}{m}{it}
  \addtoversion{bold}{\mathit}{cmr}{bx}{it}
  \newmathalphabet{\mathbfit} % math mode version of \textbfit{..}
  \addtoversion{normal}{\mathbfit}{cmr}{bx}{it}
  \addtoversion{bold}{\mathbfit}{cmr}{bx}{it}
  \newmathalphabet{\mathbfss} % math mode version of \textbfss{..}
  \addtoversion{normal}{\mathbfss}{cmss}{bx}{n}
  \addtoversion{bold}{\mathbfss}{cmss}{bx}{n}
  \ifAMStwofonts
    \ifCUPmtlplainloaded \else
      \UseAMStwoboldmath
      \makeatletter
      \new@mathgroup\upmath@group
      \define@mathgroup\mv@normal\upmath@group{eur}{m}{n}
      \define@mathgroup\mv@bold\upmath@group{eur}{b}{n}
      \edef\UPM{\hexnumber\upmath@group}
      \new@mathgroup\amsa@group
      \define@mathgroup\mv@normal\amsa@group{msa}{m}{n}
      \define@mathgroup\mv@bold\amsa@group{msa}{m}{n}
      \edef\AMSa{\hexnumber\amsa@group}
      \makeatother
      \mathchardef\upi="0\UPM19
      \mathchardef\umu="0\UPM16
      \mathchardef\upartial="0\UPM40
      \mathchardef\leqslant="3\AMSa36
      \mathchardef\geqslant="3\AMSa3E

       \let\le=\leqslant
      \let\geq=\geqslant 
    \fi
  \fi
\fi % End of NFSS release 1

\ifnfsstwo
  \DeclareMathAlphabet{\mathbfit}{OT1}{cmr}{bx}{it}
  \SetMathAlphabet\mathbfit{bold}{OT1}{cmr}{bx}{it}
  \DeclareMathAlphabet{\mathbfss}{OT1}{cmss}{bx}{n}
  \SetMathAlphabet\mathbfss{bold}{OT1}{cmss}{bx}{n}
  \ifAMStwofonts
    \ifCUPmtlplainloaded \else
      \DeclareSymbolFont{UPM}{U}{eur}{m}{n}
      \SetSymbolFont{UPM}{bold}{U}{eur}{b}{n}
      \DeclareSymbolFont{AMSa}{U}{msa}{m}{n}
      \DeclareMathSymbol{\upi}{0}{UPM}{"19}
      \DeclareMathSymbol{\umu}{0}{UPM}{"16}
      \DeclareMathSymbol{\upartial}{0}{UPM}{"40}
      \DeclareMathSymbol{\leqslant}{3}{AMSa}{"36}
      \DeclareMathSymbol{\geqslant}{3}{AMSa}{"3E}

       \let\le=\leqslant
      \let\geq=\geqslant 
    \fi
  \fi
\fi % End of NFSS release 2

\ifCUPmtlplainloaded \else
  \ifAMStwofonts \else % If no AMS fonts
    \def\upi{\pi}
    \def\umu{\mu}
    \def\upartial{\partial}
  \fi
\fi

\title{Galactic Models and White Dwarf Populations}
\author[V. Castellani et al.]
       {V. Castellani$^{1,2}$, M. Cignoni$^{1}$, S. Degl'Innocenti $^{1,2}$, S. Petroni$^1$, P.G. Prada Moroni$^{1,2}$\\
  $^1$ Dipartimento di Fisica, Universit\`a di Pisa, Via Buonarroti 2, I-56127 Pisa\\
  $^2$ Istituto Nazionale di Fisica Nucleare, Sezione di Pisa, via Livornese 1291, S. Piero a Grado, I-56100 Pisa\\
}

\date{}

\pagerange{\pageref{firstpage}--\pageref{lastpage}}
\pubyear{2002}

\begin{document}

\maketitle

\label{firstpage}

\begin{abstract}

We make use of a previous well tested Galactic model, but describing
the observational behavior of the various stellar components in terms
of suitable assumptions on their evolutionary status.  In this way we
are able to predict the expected distribution of Galactic White Dwarfs
(WDs), with results which appear in {\em rather good} agreement with recent
estimates of the local WD luminosity function.  The predicted
occurrence of WDs in deep observations of selected galactic fields is
presented, discussing the role played by WDs in star counts.  The
effects on the theoretical predictions of different White Dwarfs
evolutionary models, ages, initial mass functions and relations
between progenitor mass and WD mass are also discussed.

\end{abstract}


\section{Introduction}

Since the very beginning of modern astronomy the distribution of stars
over the night sky has been regarded as an evidence of the
distribution in space of stellar objects, i.e., in the current
terminology, as the result of the distribution of the Galactic stellar
components. However, only in relatively recent times such an evidence
has been used as an input for detailed investigations aiming to
reconstruct the spatial distribution of the various stellar
populations forming our Galaxy.  Here one may recall the ``classic''
star counts models by Gilmore \& Reid (1983, G\&R, see also e.g. Reid
\& Majewski, 1993 and Basilio et al. 1996) and by Bahcall \& Soneira
(1984, B\&S, see also e.g. Gould, Bahcall \& Maoz, 1993) as based on
the assumption of suitable spatial density distributions for the
various Galactic components and on the observational luminosity
function and colour-magnitude diagram for each stellar population.

In a recent paper (Castellani et al. 2001; Paper I) we revisited the
argument, showing that the predictions of similar Galactic models can
reach a reasonable agreement even with very deep investigations, as
recently obtained by the {\it Hubble Space Telescope} (HST). As in the
G\&R and B\&S works, that paper followed a semi-empirical approach, adopting,
as model inputs, observational constraints on the distribution of
magnitudes and colours of the various stellar populations.  However,
investigations of simple stellar populations in stellar clusters have
already repeatedly shown the good agreement between observed and
predicted CM diagrams (see, e.g., Brocato et al. 2000). This discloses
that the distribution of stars, for the evolutionary phases later than
MS, closely follows current predictions as based on theoretical
evolutionary lifetimes, independently of any assumption about the star
Initial Mass Function (IMF). In the meantime, every adopted luminosity
distribution of MS stars can be easily interpreted in terms of a
corresponding suitable behavior of the IMF. Thus, each empirical
assumption about the CM distribution of stellar populations can be
easily transferred into suitable assumptions about their original
chemical composition, IMF and age, without consequences on the
Galactic model.

The use of an evolutionary input presents several advantages.  One can
investigate the effects of varying the adopted evolutionary
parameters, constraining ages, chemical composition or IMF of the
Galactic populations. Relevant steps in such a direction have been
recently presented by several authors: Ng (1994), Ng et al. (1997) and Fan
(1999) based their simulations on population synthesis models assuming a
suitable star formation history, whereas Haywood, Robin \& Cr\'eze
(1997) (see also Robin \& Cr\'eze 1986 and Haywood 1994) based their
models on the synthesis of both the evolution of stellar population
and the dynamical evolution of the vertical structure of the disc.

However, and perhaps more interestingly, for each given IMF, chemical
composition and age, a theoretical isochrone gives rather firm
constraints on the relative abundance of stars in all the evolutionary
phases, from the MS till the final structures, either as
post-supernova objects or as cooling white dwarfs. In this paper we
will rely on a similar approach, transferring our Galactic models into
the quoted theoretical scenario, to derive theoretical
predictions concerning the Galactic distribution of white dwarf populations.

\section {The Galactic model}

As in Paper I, we will distribute the stars according to the spatial
density distribution of B\&S and G\&R Galactic models, but now relying
on suitable assumptions on the evolutionary status and on the initial
mass function (IMF) of the various Galactic populations to reproduce
the luminosity functions used as an (observational) input in Paper
I. Our model includes three components: spheroid, disc and thick
disc. The need for an extended ``thick disc'' population, as formed by
stars with spatial and kinematic properties intermediate between disc
and spheroid, was firstly suggested by Gilmore \& Reid (1983) and
further supported by detailed analysis taking into account the
velocity distribution of local stars (see e.g. Ojha et al. 1996, Wyse
\& Gilmore 1995, Norris \& Ryan 1991, Casertano, Ratnatunga \& Bahcall
1990). However the thick disc structural parameters (density law,
local density etc.)  are still debated (see e.g. Reid \& Majewski
1993, Yamagata \& Yoshii 1992, Ojha et al. 1996, Ruphy et al. 1996,
Mendez \& Guzman 1998, Buser et al. 1999, Reyl\'e \& Robin 2001).

Predicted results are obtained by randomly generating star masses
according to the adopted IMF (see Sec. 2.3) and by using stellar
models to derive luminosities in the selected bands for each given
value of the stellar mass and age.
Spheroid stars are assumed to be almost coeval and thus they are
reproduced by populating a suitable isochrone, while for both
thick disc and disc stars, one has to take into account prolonged
episodes of star formation. Thus, for these two last components, star masses
and ages are both randomly generated, the mass distribution
reproducing the selected IMF, while a flat age distribution is adopted
within the range assumed for each population.

In the following subsection we will shortly describe the main
theoretical ingredients of our model producing the distribution of the
observed nuclear burning luminous structures. The corresponding
predictions for WDs will be presented and discussed in sections 3 and 4.

\subsection{Evolutionary tracks}

The code relies on a set of homogeneous evolutionary computations
covering both the H and He burning phase for stars with original
masses in the range 0.1$\div$7 M$_{\odot}$.  Very low MS tracks are
from evolutionary calculations by Cassisi et al. (2000), which are
already shown to be in good agreement with recent theoretical
calculations, as well as with HST observations of faint MS in Galactic
Globulars (see e.g. Cassisi et al. 2000) and with recent data by Monet
et al. (1992) and Dahn et al. (1995) for subdwarfs stars in the solar
neighborhood (see e.g. Paper I). Colour transformations
are from Allard et al. (1997) model atmospheres.  However, as
discussed in Baraffe et al. (1995) and in Brocato et al. (1998),
theoretical models do not satisfactory reproduce the observed
distribution of very low mass (VLM) stars at solar metallicity. Even
if such a disagreement does not affect the WD predictions we will deal
with, to present a model as reliable as possible, for stars with
M$\le$ 0.6 M$_{\odot}$ and solar metallicity we used the empirical
V-(V-I) data by Monet et al. (1992) and Dahn et al. (1995) and the
observational LF by Wielen, Jahreiss \& Kruger (1983).  
For M$>$0.6M$_{\odot}$ we use the Cassisi et al. (1998) and the
Castellani, Degl'Innocenti \& Marconi (1999) evolutionary tracks up to
the end of the AGB, which are in excellent agreement with recent
results from the Hipparcos satellite for nearby stars (Kovalevsky
1998) and open clusters (see e.g. Castellani, Degl'Innocenti \& Prada
Moroni 2001, Petroni 1999) and with observational data of globular
clusters at different metallicities (see e.g. Cassisi et al. 1999,
Brocato et al. 2000). The adopted colour transformations are from
Castelli et al. (1997).

\subsection{IMF}

Several authors (see e.g. Scalo 1998, Kroupa 2000) discussed in
details local star counts in order to infer the underlying IMF.
%Although many researchers have measured the IMF in a 
%variety of Galactic regions a firm conclusion is still not available.  
While the most of authors agree on a Salpeter IMF for M$\ga$ 0.5
M$_{\odot}$ (see e.g. Massey et al. 1995, Kroupa 2001a) the IMF
behaviour for lower masses is still uncertain (see e.g. Scalo 1998,
Kroupa 2001a, Reyl\'e \& Robin 2001). The uncertainty on the IMF for
low mass stars reflects the observational errors on the local
luminosity function (LF) at low luminosity for disc and spheroid stars
(see e.g. Gould, Flynn \& Bahcall 1998, Paper I, for a
discussion).  Moreover, some authors (Scalo 1998, Kroupa 2001b,
Eisenhauer 2001) suggest that the IMF could vary with time and with
environment, an hypothesis which is rejected by other researchers (see
e.g. Gilmore 2001).

In our model we assume the ``average Galactic field IMF'' proposed by
Kroupa (2001b) for the three Galactic components, that is dN/dM
$\propto$ M$^{-\alpha}$ with :\\ $\alpha = 1.3 (\pm 0.5)$ \hspace{1cm}for
\hspace{1cm} 0.08 $\le$ m/M$_{\odot} < 0.5$ \\ $\alpha = 2.3 (\pm 0.3)$
\hspace{1cm}for \hspace{1cm} 0.50 $\le$ m/M$_{\odot}$ \\ As a whole,
with the adopted IMF and evolutionary tracks, theoretical luminosity
functions appear in agreement with the observed ones within their
observational uncertainties.

\subsection{Spheroid}

As usually, we adopt for the spheroid the de Vaucouleurs density
law. The assumed Galactocentric distance is R$_{\rm o}$=8 Kpc and the axis
ratio for the spheroid is 0.8
%Incidentally we notice that the extimated flattening from recent analysis ranges
%between 0.6 to 1.0 with some indication for a flattening around 0.8 
(see Robin, Reyl\'e \& Cr\'eze 2000, for a discussion).  
{\em Following Gilmore \& Reid (1983, see also e.g. Basilio et al. 1996)
we adopt the halo/disc density ratio of 0.125\%).}
A theoretical
isochrone of 12 Gyr (Z=0.0002, Y=0.23),
populated by using the Kroupa (2001a) IMF, is assumed. 
The chosen theoretical approach is
spontaneously producing predictions about the occurrence of stars in
the various evolutionary phases. As an example, Fig. 1
shows the predicted CM diagram on a selected area of 1 square degree
at the North Galactic Pole (NGP) with the contribution by
subgiant (SGB) and red giant (RGB) stars. The only one HB star
predicted in that field is not plotted, since no firm theoretical
constraints exist for its colour.

%==============================================  FIGURA 1

\begin{figure}
\label{soloalone}
\centerline{\epsfxsize= 8 cm \epsfbox{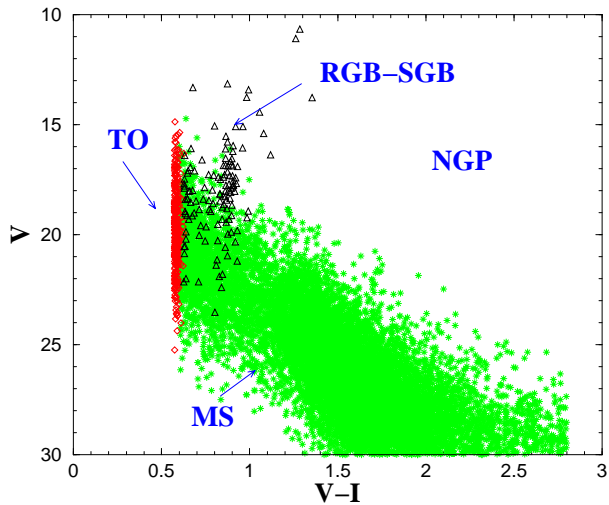}} 
\caption{Theoretical CMD for spheroid stars in a field of area 1 square
degree at the North Galactic Pole (NGP).  Different symbols represent
stars in the various evolutionary phases.}
\end{figure}
%================================================

\subsection{Disc}

The adopted disc density law is the widely used double exponential in
the Galactocentric distance on the Galaxy plane and in height over the
plane, with scale length and scale height of 3500 pc and 325 pc,
respectively (Reid \& Majewski 1993, Basilio et al. 1996). We assume
a constant star formation rate (SFR) from 50 Myr to 9 Gyr to populate
evolutionary tracks  with Z=0.02 Y=0.27.

\subsection{Thick disc}

The thick disc density law can be reasonably modeled either by a
double exponential or by a density law close to sech$^{2}$(z). Star
counts are unable to distinguish between these two hypotheses (Reyl\'e
\& Robin 2001) and the thick disc structure is generally described as
a double exponential, with horizontal and vertical scales
approximately in the ranges r$_0$$\approx$2.5$\div$4.0 Kpc,
z$_0$$\approx$600$\div$1600 pc, respectively.  The scale length and
height and the thick disc/disc density ratio suggested by Gilmore \&
Reid (1983) (see also e.g. Basilio et al. 1996) and adopted in this
paper are 3500 pc, 1300 pc and 2\%, respectively.  Following Gilmore,
Wyse \& Jones (1995) and Norris (1999) we choose for the thick disc a
metallicity of $\sim$Z=0.006 and a SFR centered at $\sim$ 10 Gyr with
a spread of few Gyr.

\subsection{Reddening/obscuration}

We adopt for intermediate-to-high Galactic latitudes
($|\mathrm b|\geq$10$^{\circ}$) the E(B-V) reddening maps by Burnstein
\& Heiles (1982). The visual extinction A$_{\mathrm V}$=3.12 E(B-V) and the
reddening in the infrared colours are from Bessell \& Brett (1988, see
also Clementini et al. 1995). A reddening scale height of 100 pc is
adopted (see e.g. Mendez \& van Altena, 1998).

\section{The white dwarf population}

According to the discussion given in Paper I, the Galactic model
presented in the previous section appears able to reasonably account
for available star counts, even down to very faint magnitudes
(V$\approx 27$). By relying on this model one can easily predict the
expected abundance of WDs, since each star evolved beyond the AGB
phase and less massive than the lower mass limit for supernovae
(M$_{\rm up}$) is assumed to be a WD. However, to predict the CM
location of white dwarfs one needs further theoretical ingredients, as
given by: i) a WD mass - progenitor mass relation ii) theoretical WD
models giving luminosity and temperature of a WD as a function of mass
and age and iii) suitable colour transformations.  The WD cooling age
(that is the time spent on the cooling curve) is simply given by the
difference between the age associated to the star and the age of the
WD progenitor at the end of the AGB.  For T$_{eff} < 4000 ^{\mathrm
o}$K we adopt the colour relations by Saumon \& Jacobson (1999), which
include a detailed treatment of collision induced absorption of H$_2$,
whereas for higher temperature, the results of Bergeron, Wesemael \&
Beauchamp (1995) were used.  As an example Fig. 2 shows
theoretical WDs isochrones.

%==============================================  FIGURA 2

\begin{figure}
\label{isonane}
\centerline{\epsfxsize= 8 cm \epsfbox{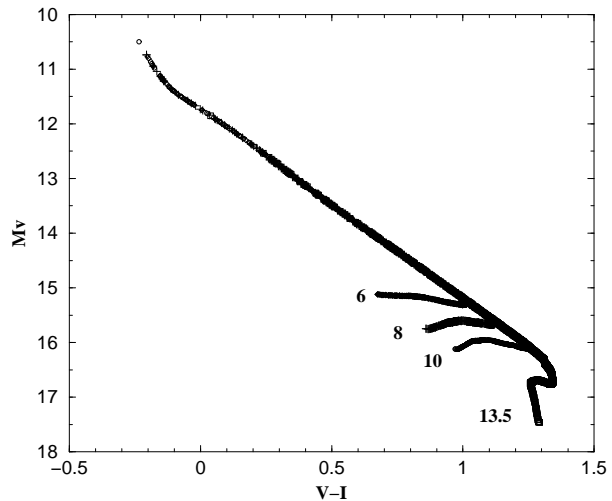}} 
\caption{Theoretical isochrones in the $M_V-(V-I)$ diagram for
spheroid WDs of the labeled ages (in Gyr).}
\end{figure}
%================================================

{\em We notice that in our model we include only WDs with hydrogen
atmosphere (DA WDs) which are shown to be the most numerous. However,
the last two bins of the observed disc WD LF by Legget, Ruiz \&
Bergeron (1998, see Fig. 7) are populated by a relevant number of
non-DA WDs. As well known, it is still not clear if this phenomenon is
due to a chemical evolution of the WD outer layers or to the relevant
differences in the cooling times between DA and non-DA.
Thus, given the present
uncertainty in the theoretical scenario we prefer, as first step,
do not include non-DA WDs.}

\begin{figure}
\centerline{\epsfxsize= 7 cm \epsfbox{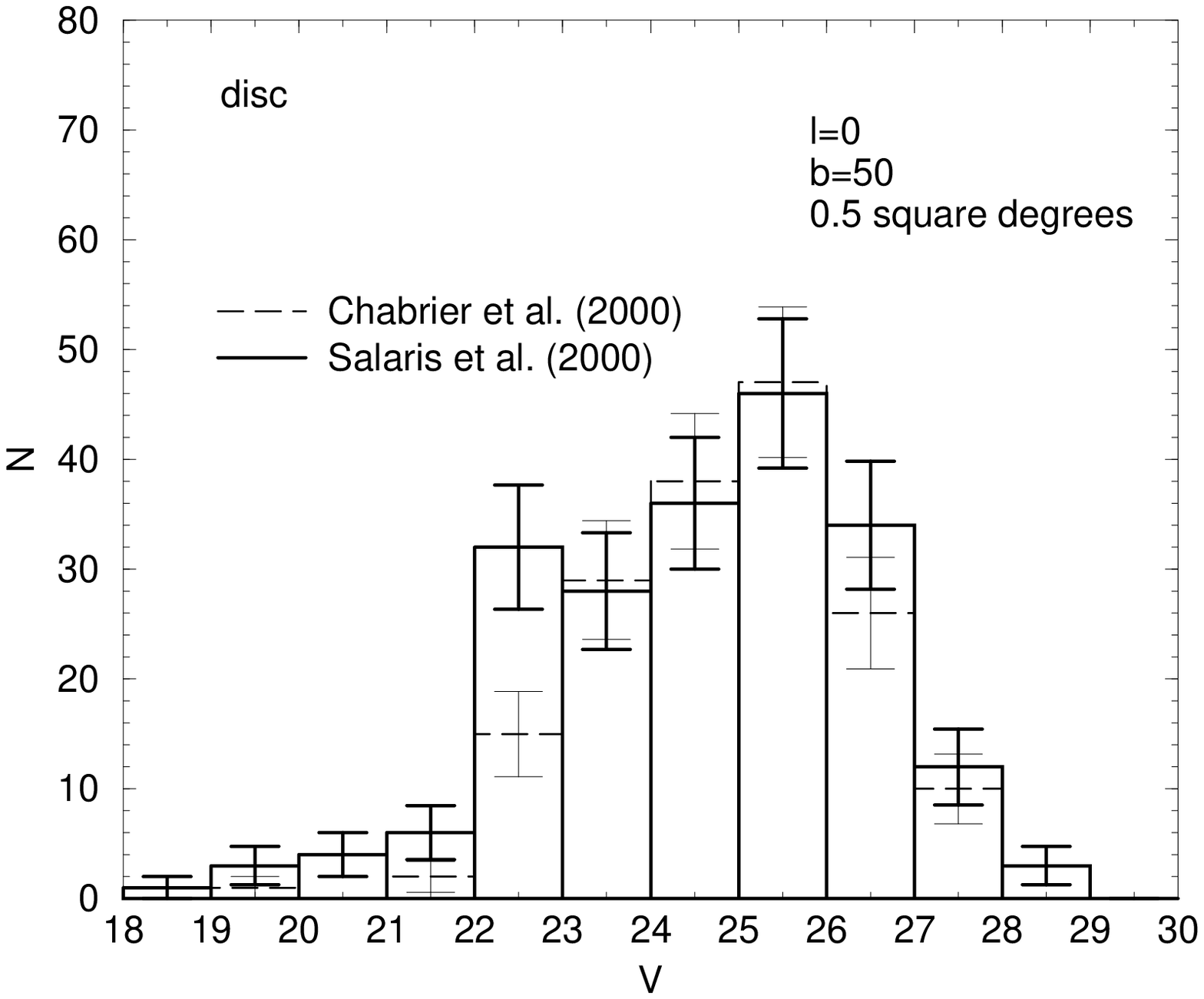}} 
\centerline{\epsfxsize= 7 cm \epsfbox{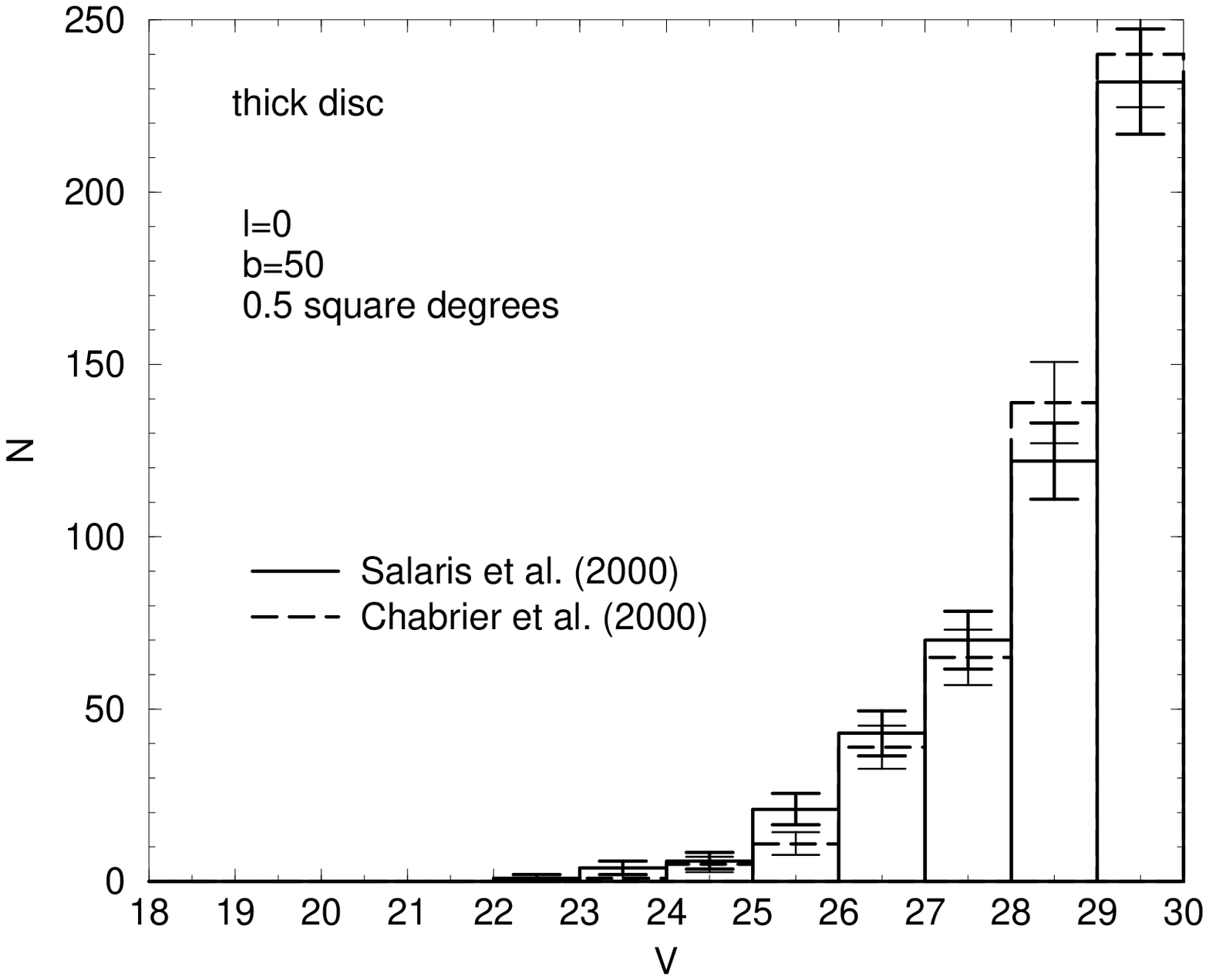}} 
\centerline{\epsfxsize= 7 cm \epsfbox{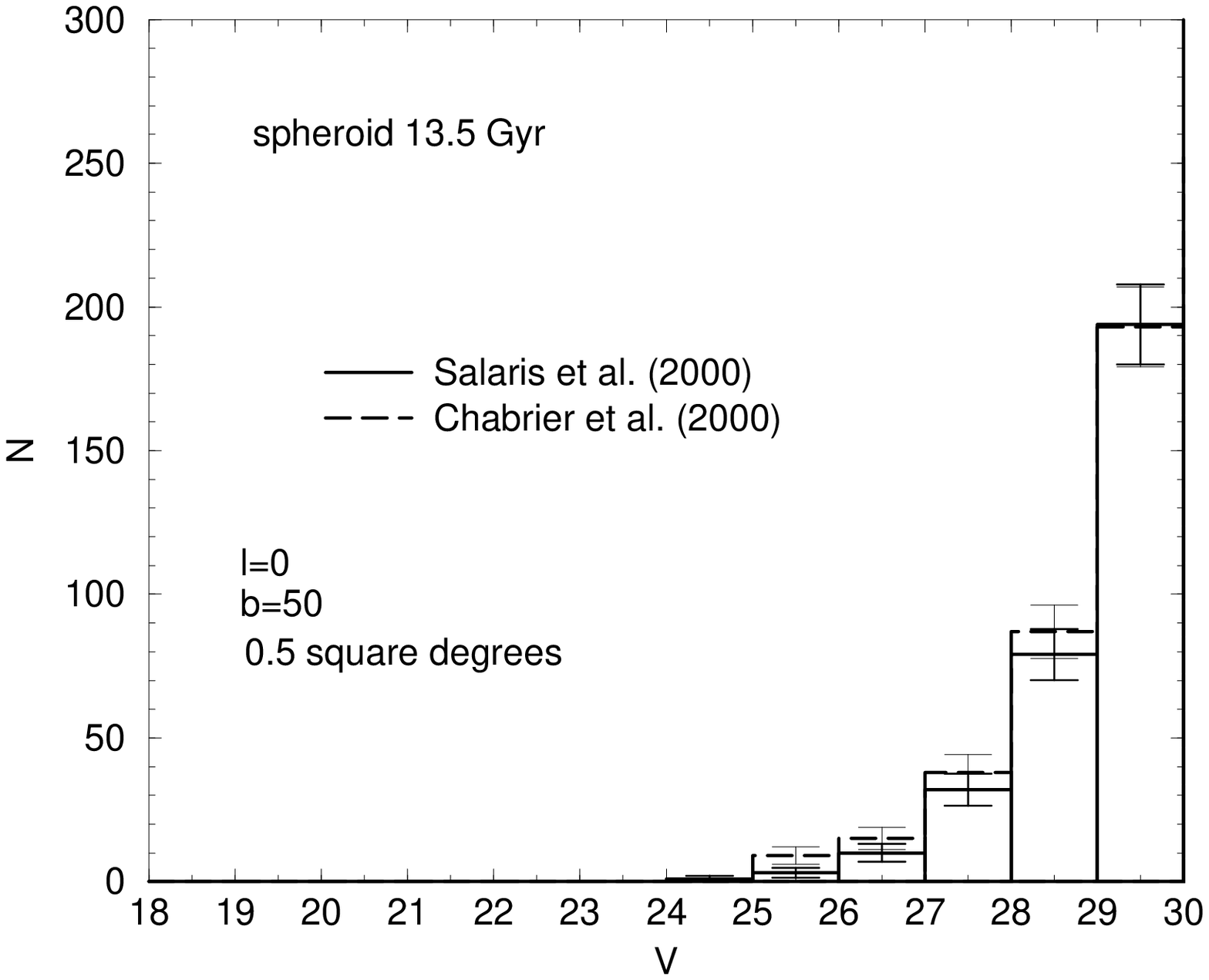}} 
\caption{Comparison between the predicted V-magnitude distribution (in
a field at the Galactic longitude $l={0}^{\circ}$ and at the Galactic
latitude $b={50}^{\circ}$, of extension 0.5 square degrees) obtained
by adopting theoretical WD cooling tracks by Chabrier et al. (2000)
(solid line) and by Salaris et al. (2000) (dashed line) for the disc
(upper panel), the thick disc (middle panel) and the halo (lower
panel). The error bars indicate the poissonian statistical uncertainty on
the counts. The WD mass - progenitor mass relation is by Weidemann
(2000) for the disc/thick disc and by Dominguez et al. (1999) for the
spheroid.}
\label{tracce}
\end{figure}

The theoretical prediction of WD populations is affected by several
uncertainties. First of all, theoretical models are not settled yet,
as shown by the somewhat large differences among recent models in the
literature (Wood 1995, Benvenuto \& Althaus 1999, Hansen 1999,
Salaris et al. 2000, Chabrier et al. 2000, 
Castellani, Prada Moroni \& Straniero 2001).  
To investigate the dependence of the WD distribution on
the adopted cooling sequences, Fig. \ref{tracce} shows predicted results
with two different sets of theoretical WD models, as given by Chabrier
et al. (2000) and Salaris et al. (2000), which are the only models
cooling down until an age of the order of the age of the Universe for
a suitable range of masses. 
%==============================================  FIGURA 3
%==============================================
Figure \ref{tracce} compares the magnitude distribution for spheroid,
disc and thick disc WDs for the two sets of WD tracks; luckily enough,
one finds that the differences are within the statistical
uncertainties.

\begin{figure}
\centerline{\epsfxsize= 8 cm \epsfbox{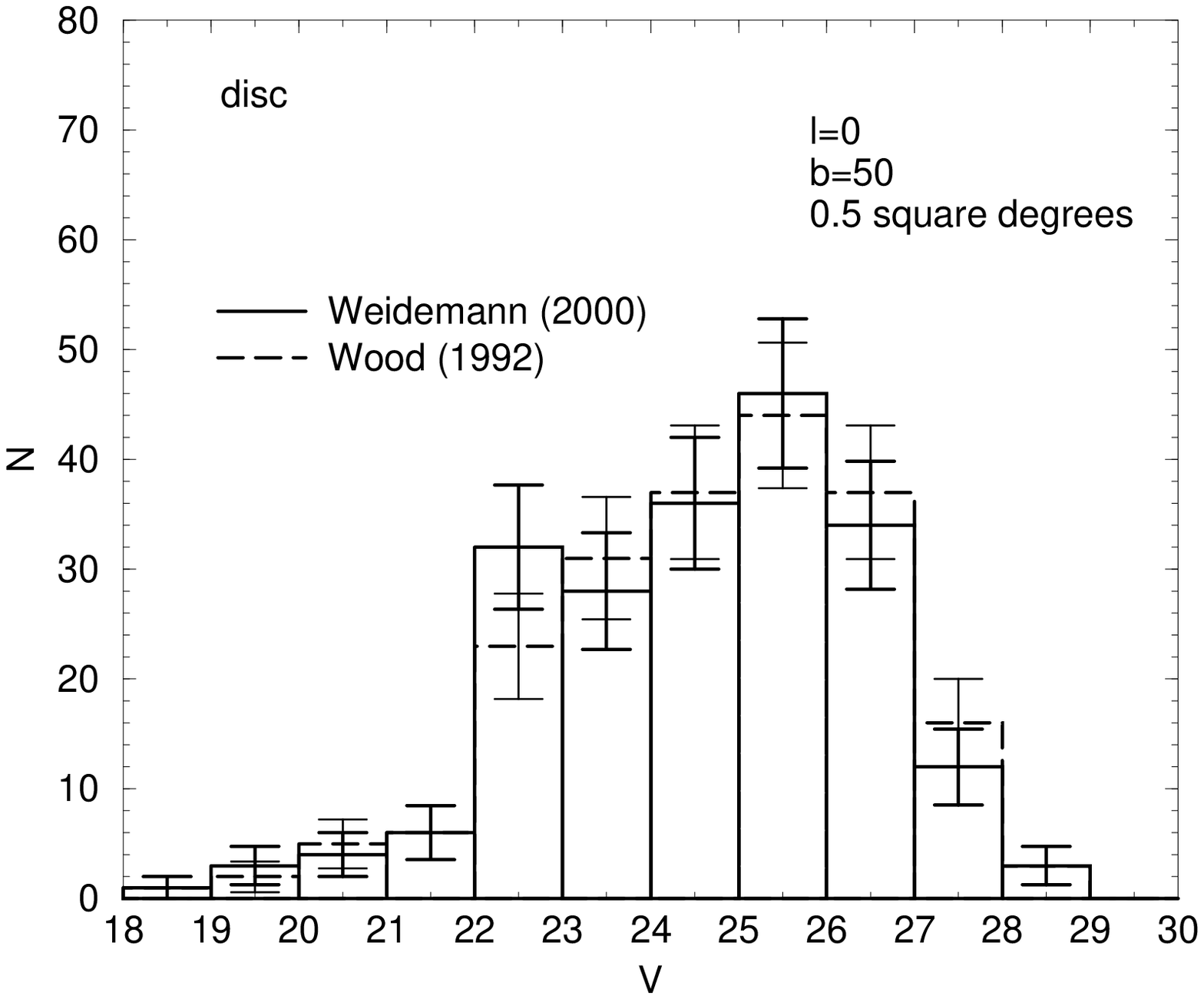}}
\centerline{\epsfxsize= 8 cm \epsfbox{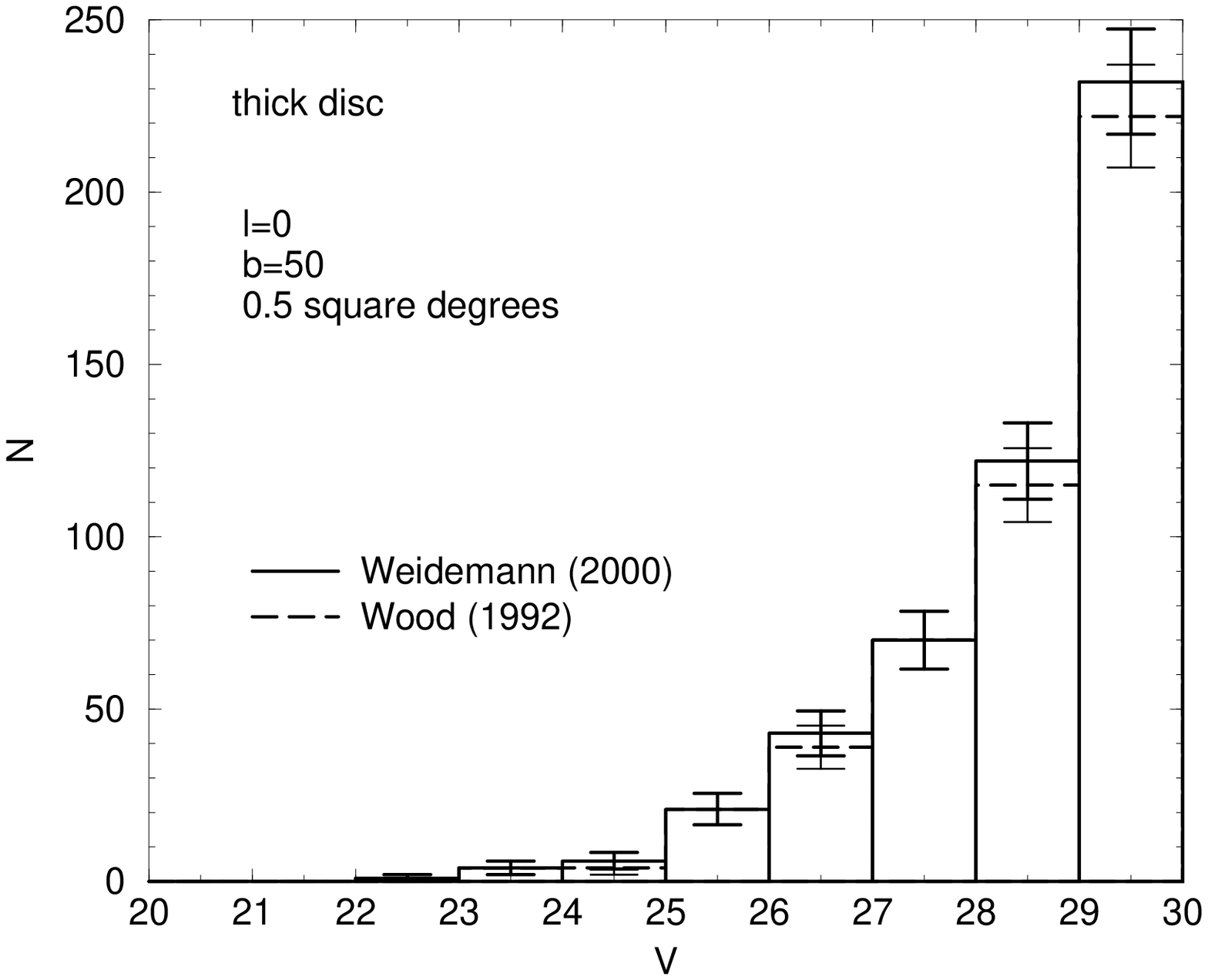}} 
\caption{Comparison between the predicted V-magnitude distribution (in
a field at $l={0}^{\circ}$, $b={50}^{\circ}$ of extension 0.5 square
degrees) as obtained by adopting the initial-final mass relation by
Wood (1992) (solid line) and by Weidemann (2000) (dashed line) for the
disc (upper panel) and the thick disc (lower panel). The adopted WD
tracks are the Salaris et al. (2000) ones.  The error bars indicate
the poissonian statistical uncertainty on the counts.}
\label{mi-mf}
\end{figure}

Another cause of uncertainty is the adopted relation between the WD
mass and the progenitor mass. As well known, this indetermination is
connected to the uncertainties in the final evolution of AGB stars,
particularly during the thermal pulses and at the onset of the
super-winds, and to the still present debate about the predicted
extension of the convective core during central burning phases (see
e.g. Dominguez et al. 1999).  For the spheroid we adopt the theoretical
relation by Dominguez et al. (1999), where the final WD mass is given
by the helium core mass at the first thermal pulse, as obtained by
assuming a standard extension of the convective core. For disc and
thick disc one has the alternative choice between the semi-empirical
relation by Weidemann (2000), as inferred from the comparison between
observations of young open cluster and theoretical models and the
exponential one by Wood (1992), based on the Planetary Nebula Nuclei
mass distribution.  Comparison between the predicted WD
magnitude distribution, as given in Fig. \ref{mi-mf}, under
the two alternative hypothesis, again discloses relatively small
differences. Thus, we made the choice of using the theoretical relation
by Weidemann (2000).

%%%%%%%%   FIGURA 4

%%%%%%%%%%%%%%%%%%%%%%%%%%%%%%%%%

\subsection{The population parameters}

\begin{figure}
\centerline{\epsfxsize= 8 cm \epsfbox{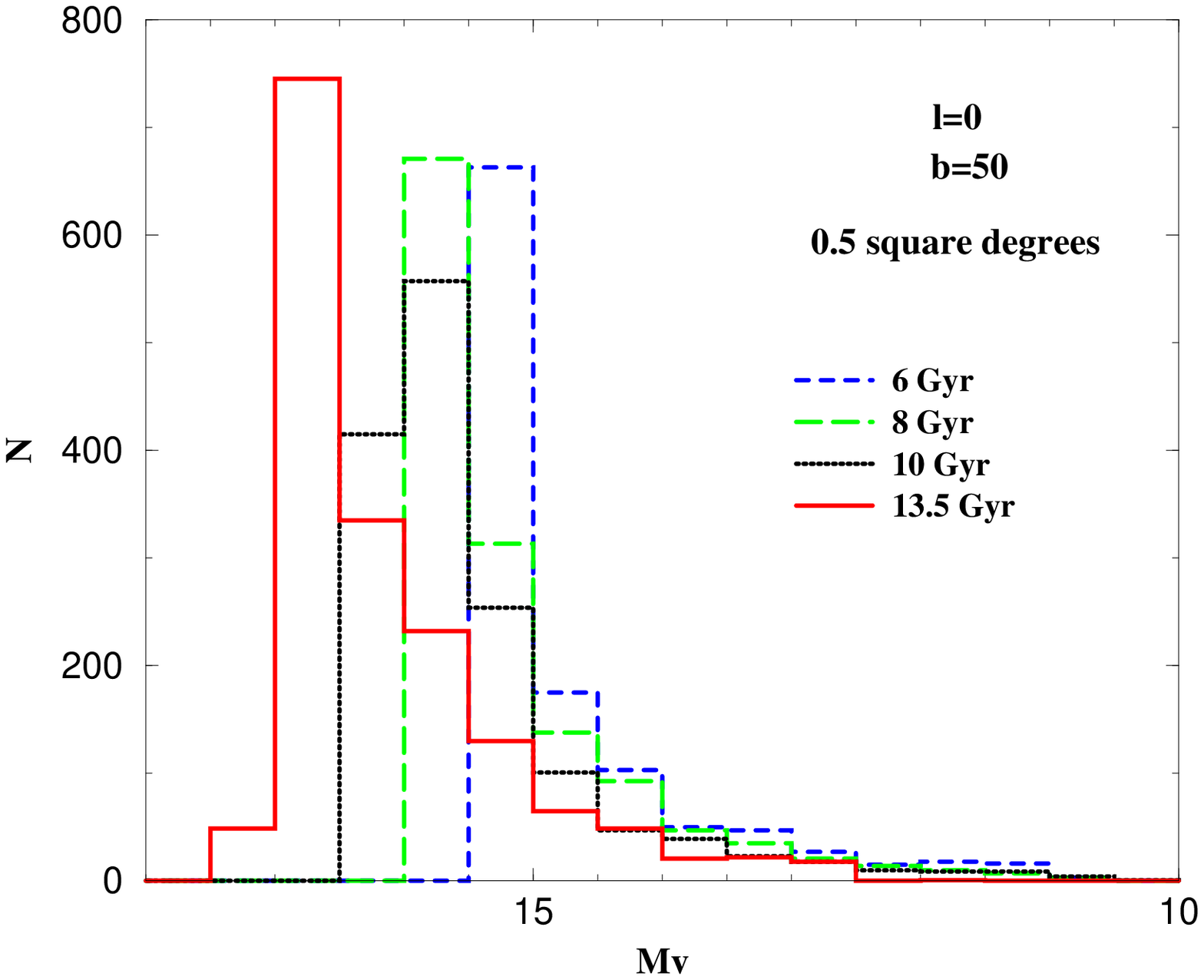}}
\centerline{\epsfxsize= 8 cm \epsfbox{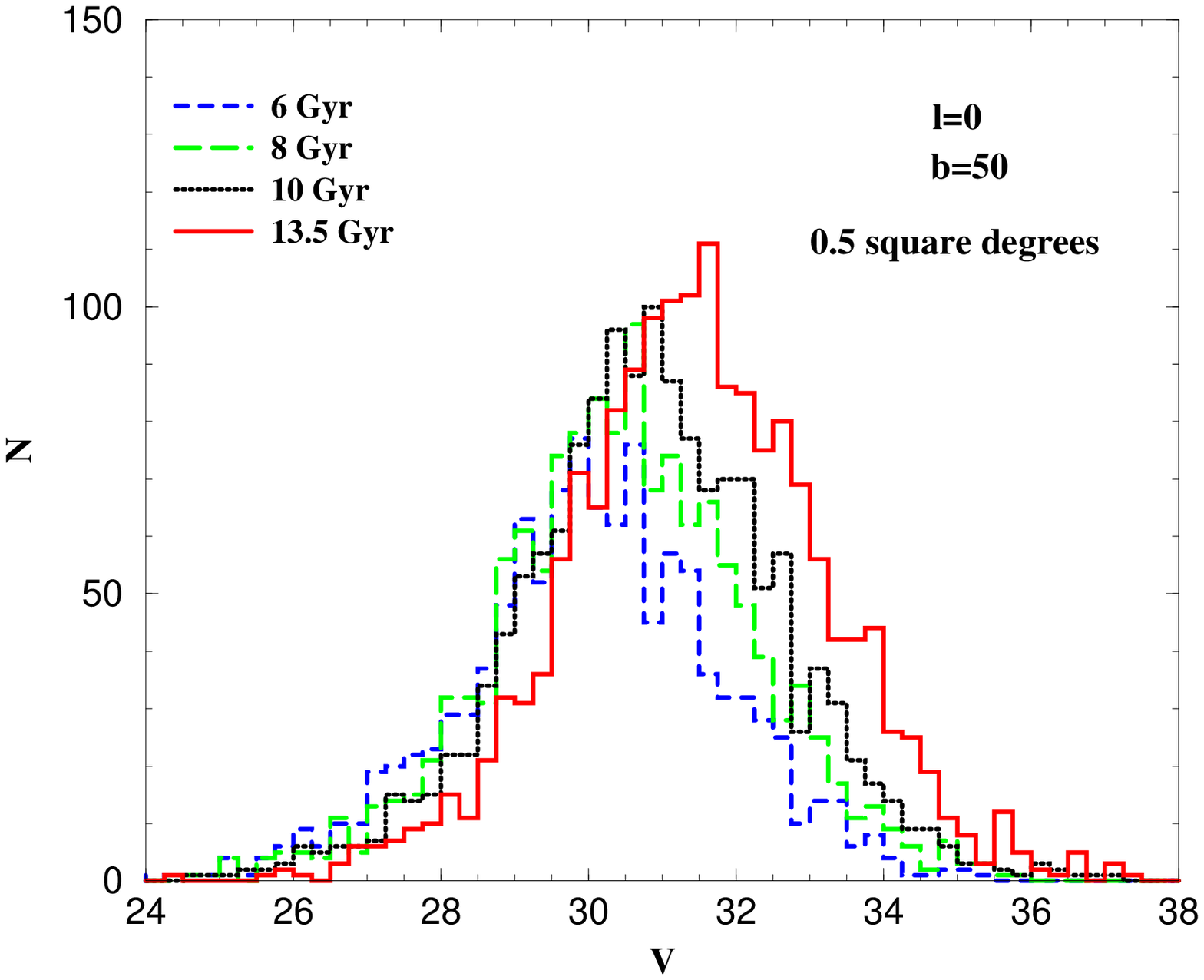}}
\caption{The predicted luminosity function (upper panel) and
V-magnitude distribution (lower panel) for spheroid WDs at different
ages (in a field at $l={0}^{\circ}$, $b={50}^{\circ}$ of extension 0.5
square degrees). The adopted WD cooling tracks are from Salaris et
al. (2000). The Dominguez et al. (1999) theoretical relation between
the progenitor mass and the WD mass is used.}
\label{WD_age}
\end{figure}

Before presenting more detailed predictions, and to give light on the
possible variations of the adopted theoretical scenario, let us give a
short discussion on the influence of the input parameters concerning
the stellar population.  As a first point the predicted halo WD
population obviously depends on the assumption on the halo age.
Figure \ref{WD_age} shows the predicted luminosity function at four
different ages: 6, 8, 10 and 13.5 Gyrs.  As well known (see e.g. Brocato
et al. 1999), for a given age, the bulk of the WD distribution is
close but not precisely at the faint end of the cooling isochrone. The
larger is the age the fainter is the bulk of the WD population.  This
is a well understood feature: the increase of the time spent in the
cooling sequence implies a progressive decrease of WD luminosity.
Figure \ref{WD_age} shows the related theoretical predictions.

%%%%%%%%%%   FIGURA 5

%%%%%%%%%%%%%%%%%%%%%%%%%%%%%%

\begin{figure}
\centerline{\epsfxsize= 7 cm \epsfbox{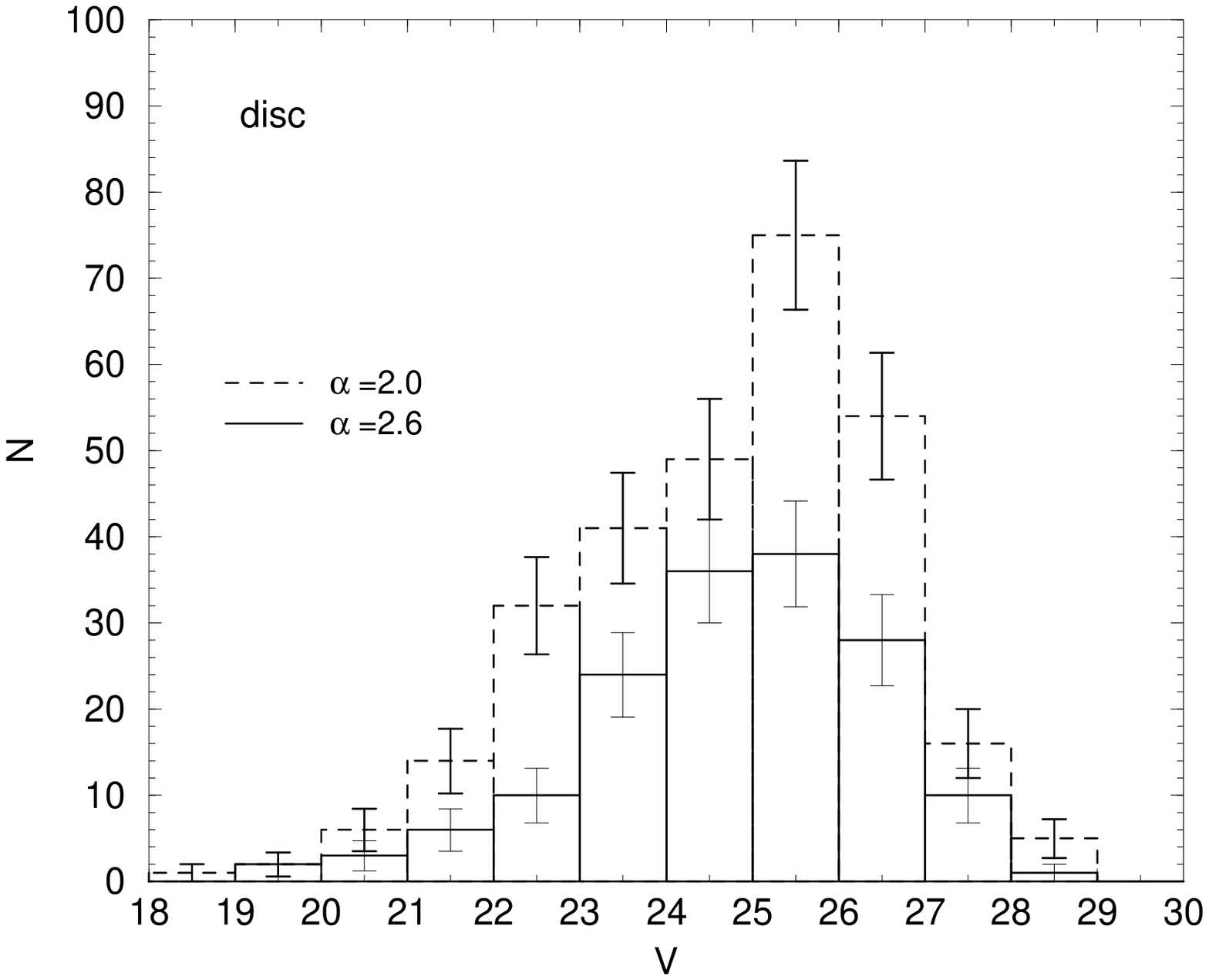}}
\centerline{\epsfxsize= 7 cm \epsfbox{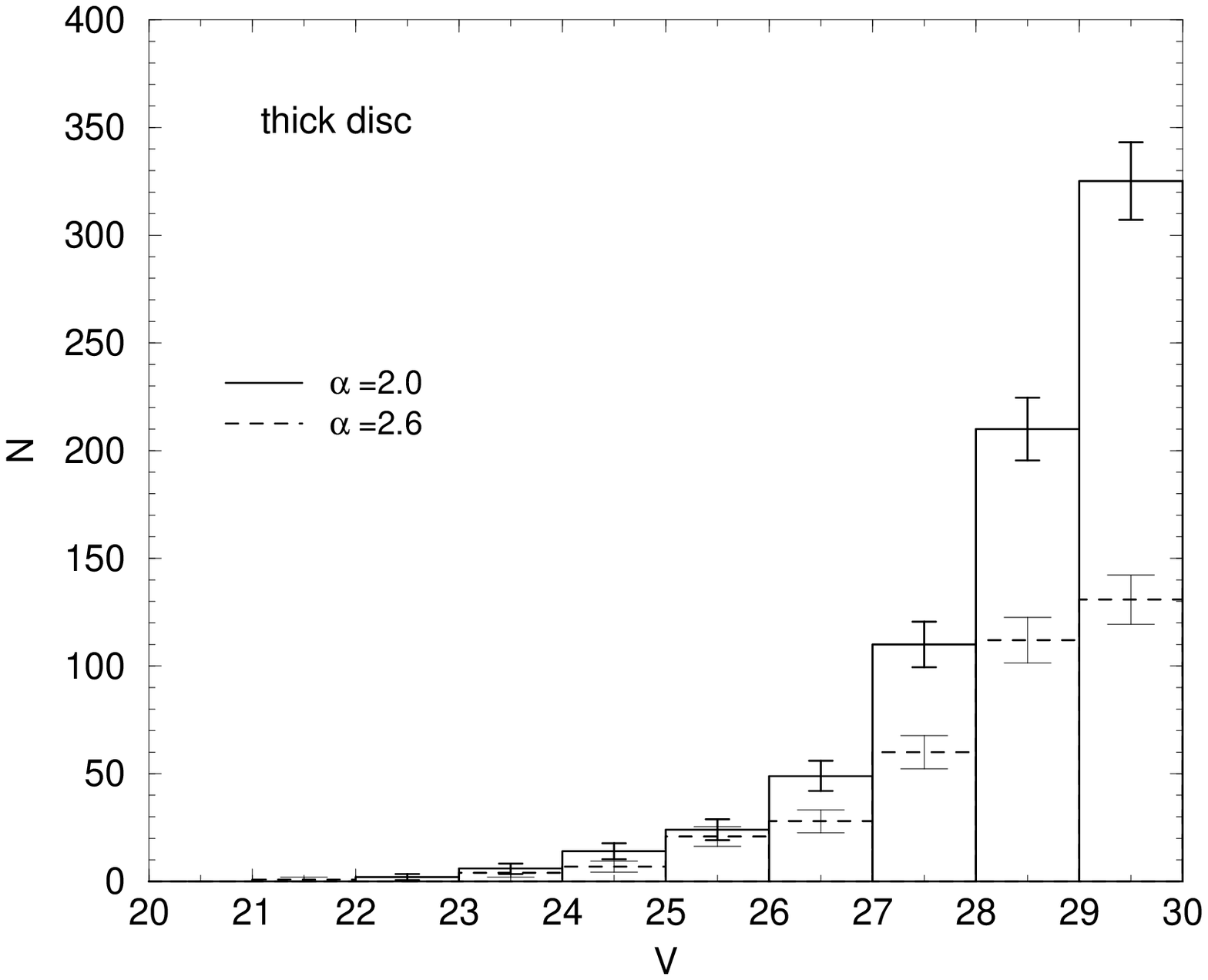}}
\centerline{\epsfxsize= 7 cm \epsfbox{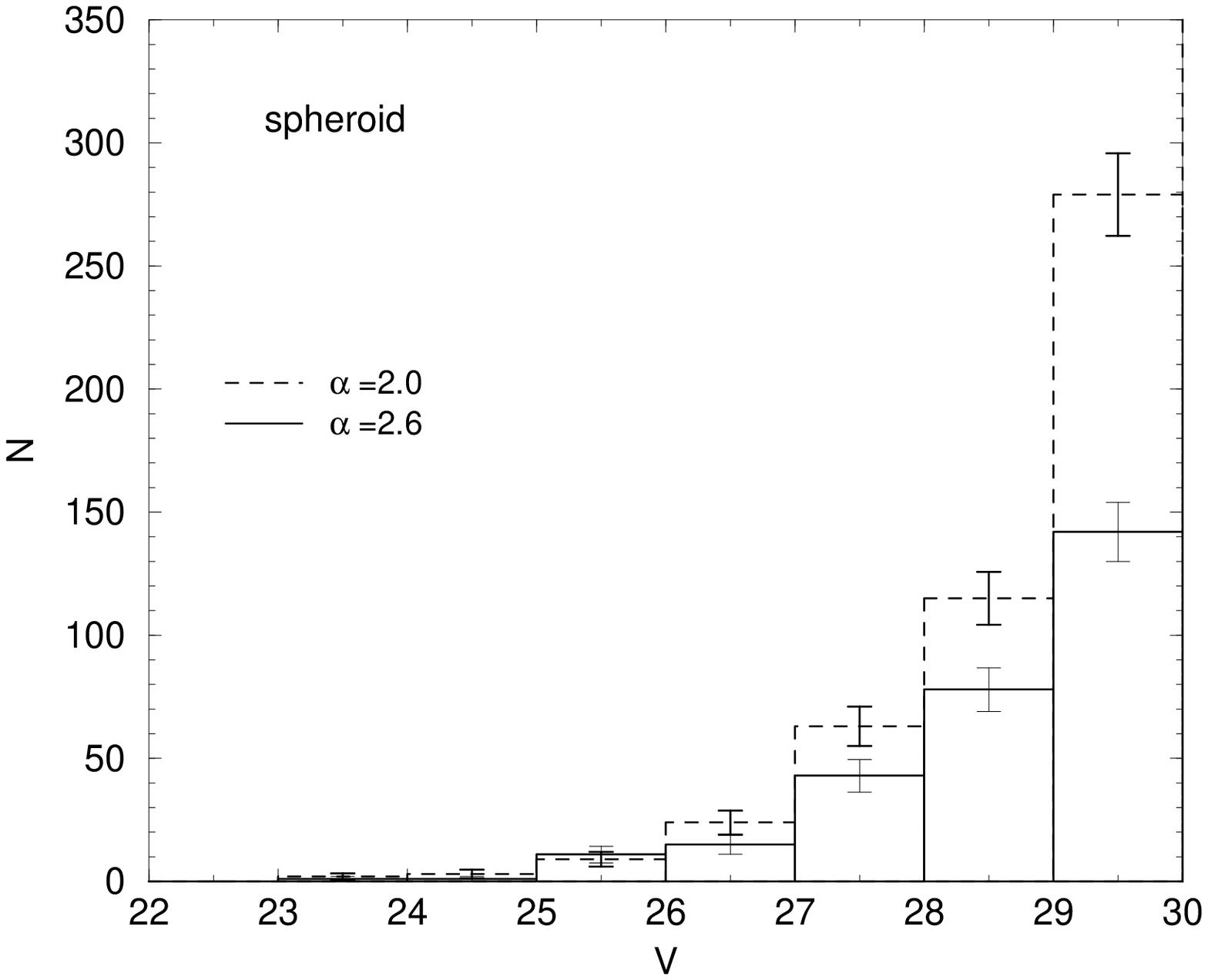}}
\caption{Comparison between the predicted V-magnitude distribution (in
a field at $l={0}^{\circ}$, $b={50}^{\circ}$ of extension 0.5 square
degrees) obtained by adopting, for $M>0.6M_{\odot}$, as IMF exponent
either $\alpha=2.0$ (solid line) or $\alpha=2.6$ (dashed line) for the
disc (upper panel), the thick disc (middle panel) and the spheroid
(lower panel). The error bars indicate the poissonian statistical
uncertainty on the counts. The adopted theoretical WD tracks are from
Salaris et al. (2000). The assumed age for the spheroid is 13.5 Gyr.}
\label{WD_IMF}
\end{figure}

Furthermore, the distribution of WD populations significantly depends
on the adopted IMF. The number of WDs is obviously affected only by
IMF variations for masses which could evolve into WDs in a time
shorter than the estimated age of the Universe. Figure \ref{WD_IMF}
shows the effect on the distribution of a variation of the IMF for
masses greater than 0.6 M$_{\odot}$.
%%%%%%%%%%%%%%   FIGURA 6
%%%%%%%%%%%%%%%%%%%%%%%%%%%%
As range of variation we assume the uncertainty on the Salpeter
exponential, as evaluated by Kroupa (2001a); see Sec. 2.2.  As
expected, a steeper IMF ($\alpha$=2.6) depopulates the WD stars. This
behaviour can be easily understood as a consequence of the decrease of
the number of stars in the mass range able to produce WDs.

\section{Results}

As a first test of the model, Fig. 7 compares the predicted local WD
luminosity function with recent observations by Liebert, Dahn \& Monet
(1988) and Leggett et al. (1998); {\em the agreement appears good, even if
it is not perfect.}  We remind that the only normalization adopted in
our model is the one of hydrogen-burning stars in the solar
neighborhood.  Thus, the density distribution of WDs naturally arises
from the model and the plotted WD luminosity function is just the
output of our code without any additional normalization.  {\em To
reproduce the local white dwarfs luminosity function, we restricted
our calculations to distances from the Sun up to 200 pc, as
obtainable from Table 2 of Legget et al. (1998).  The histogram of
Fig.7 does not include thick disc white dwarfs because we checked
that, due to the very small distance from the Sun, their contribution is
negligible. Comparisons between theoretical and observed WD LFs 
has been presented by Legget et al. themselves by
adopting WD evolutionary sequences by Wood (1995). Their
fit appears similar to our results. We remind that our LF
directly arises from a Galactic model built with a given spatial
distribution and defined parameters, so that we do not apply any normalization to
the WD LF itself. On the contrary, the Legget et al. LF is obtained only
from WD isochrones (without any assumption
on the WD spatial distribution) and then it is normalized to the observational data. 
A detailed analysis of disc WD LFs has been presented by Hansen (1999) by using
his WD isochrones. The author reproduced very well the LF profile showing the
well known influence of the disc age on the cutoff region, even in this case no 
assumptions have been made on the Galactic spatial distributions.
Also Hansen found that the inclusion of a given percentage of non-DA WDs can smooth the LF
profile. It is thus possible that we could reach an even better
agreement with observations by slightly changing the assumed disc age
or by including non-DA WDs. Moreover even a variation of the model
parameters can slightly modify the results,  however a  deep analysis of
this problem is out of the purposes of the present paper.} 

By relying on the obtained agreement between theoretical and observational WD LFs, in
this section will present the predictions by our model for three
fields in different directions of the Galaxy.

%=====================FIGURE 7=============================================
\begin{figure}
\label{HRvr}
\centerline{\epsfxsize= 8 cm \epsfbox{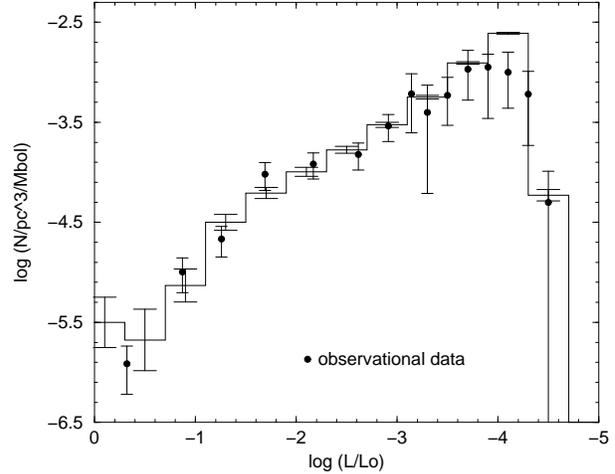}}
\caption{Comparison between the predicted local WD luminosity function 
 and recent observations by Liebert, Dahn \& Monet (1988) and
Leggett, Ruiz \& Bergeron (1998). {\em Statistical (poissonian) errors in the theoretical histogram
are also shown (thin lines).}}
\end{figure}
%===========================================================================
%An age of 12
%Gyrs is assumed for the spheroid population.  We notice that this
%paper is focused on presenting our theoretical Galactic model.  
Future comparisons with observations will be fundamental to infer
firmer constraints on the parameters assumed in the model whose values
is not definitively established yet.

The $(V-I, V)$ colour-magnitude diagrams (CMDs) for the field stars at
the North Galactic Pole, at the Galactic coordinates $l={0}^{\circ}$,
$b={50}^{\circ}$ and $l={0}^{\circ}$, $b={30}^{\circ}$ are shown in
Fig. 8; the extension of each field is 0.5 square degrees.  The CMDs
are extended down to magnitude $V=38$ to display the strong presence
of spheroid WDs at extremely low luminosities.
%==============================================  FIGURA 8
\begin{figure}
\label{HR}
\centerline{\epsfxsize= 7 cm \epsfbox{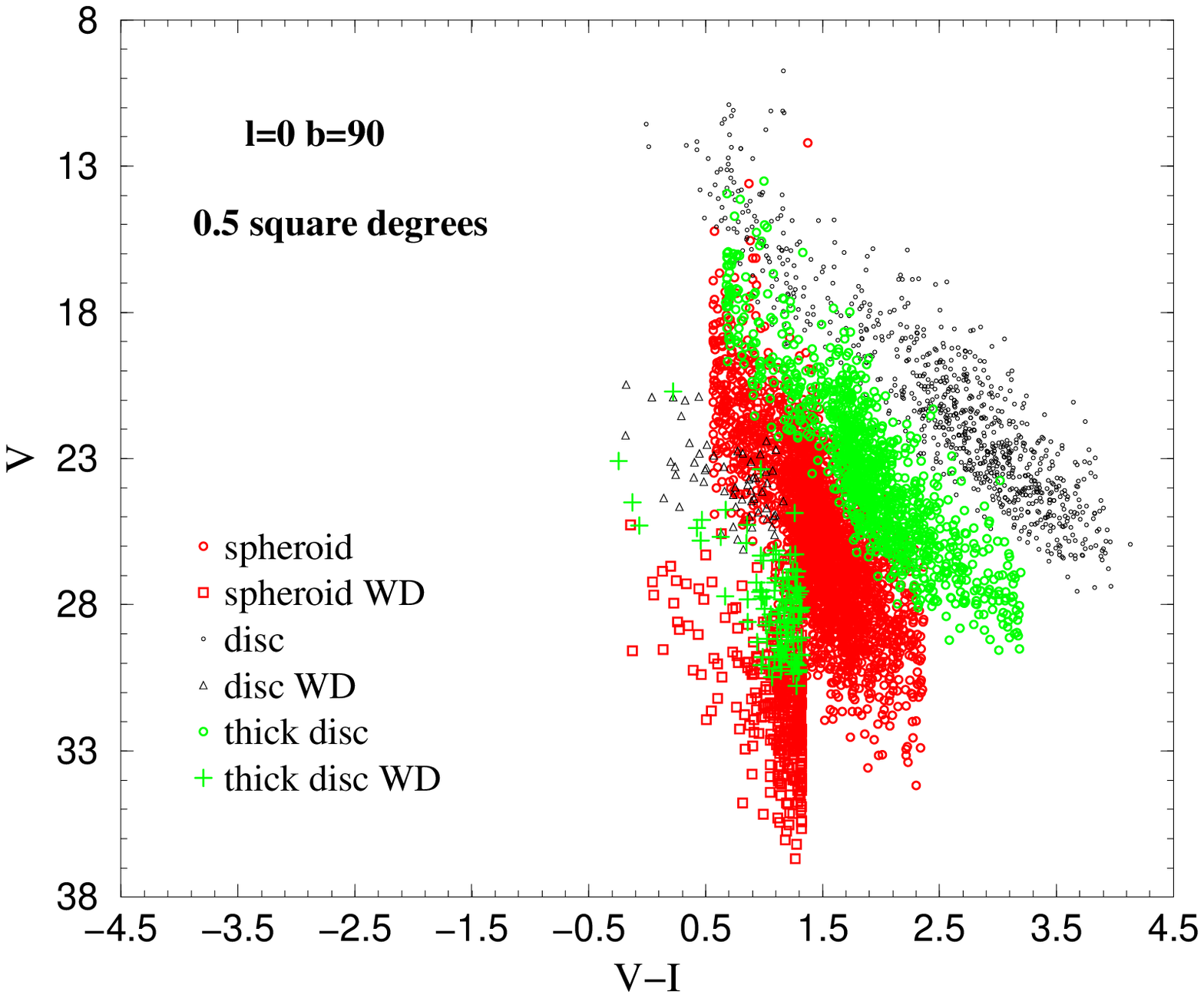}} 
\centerline{\epsfxsize= 7 cm \epsfbox{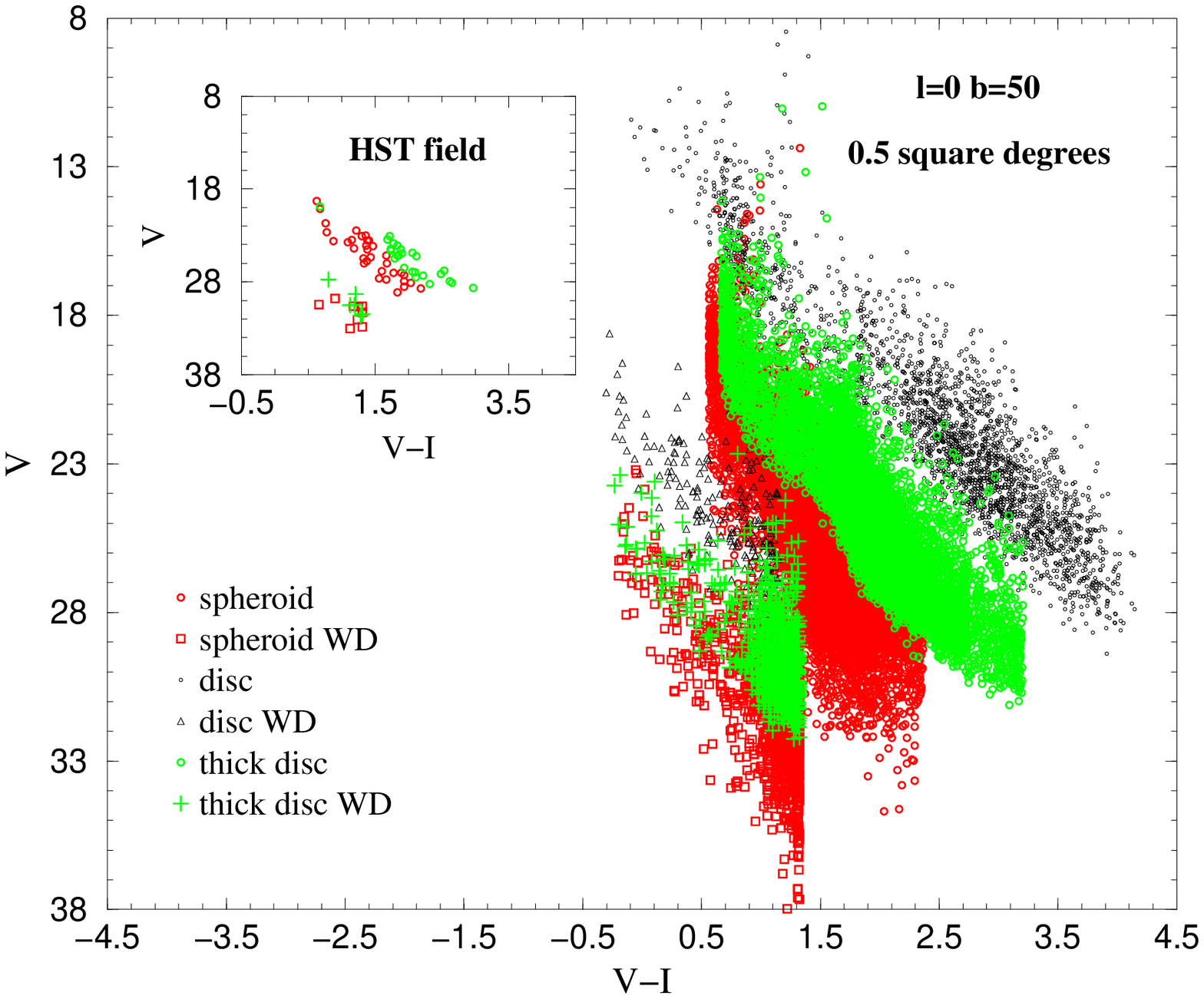}}
\centerline{\epsfxsize= 7 cm \epsfbox{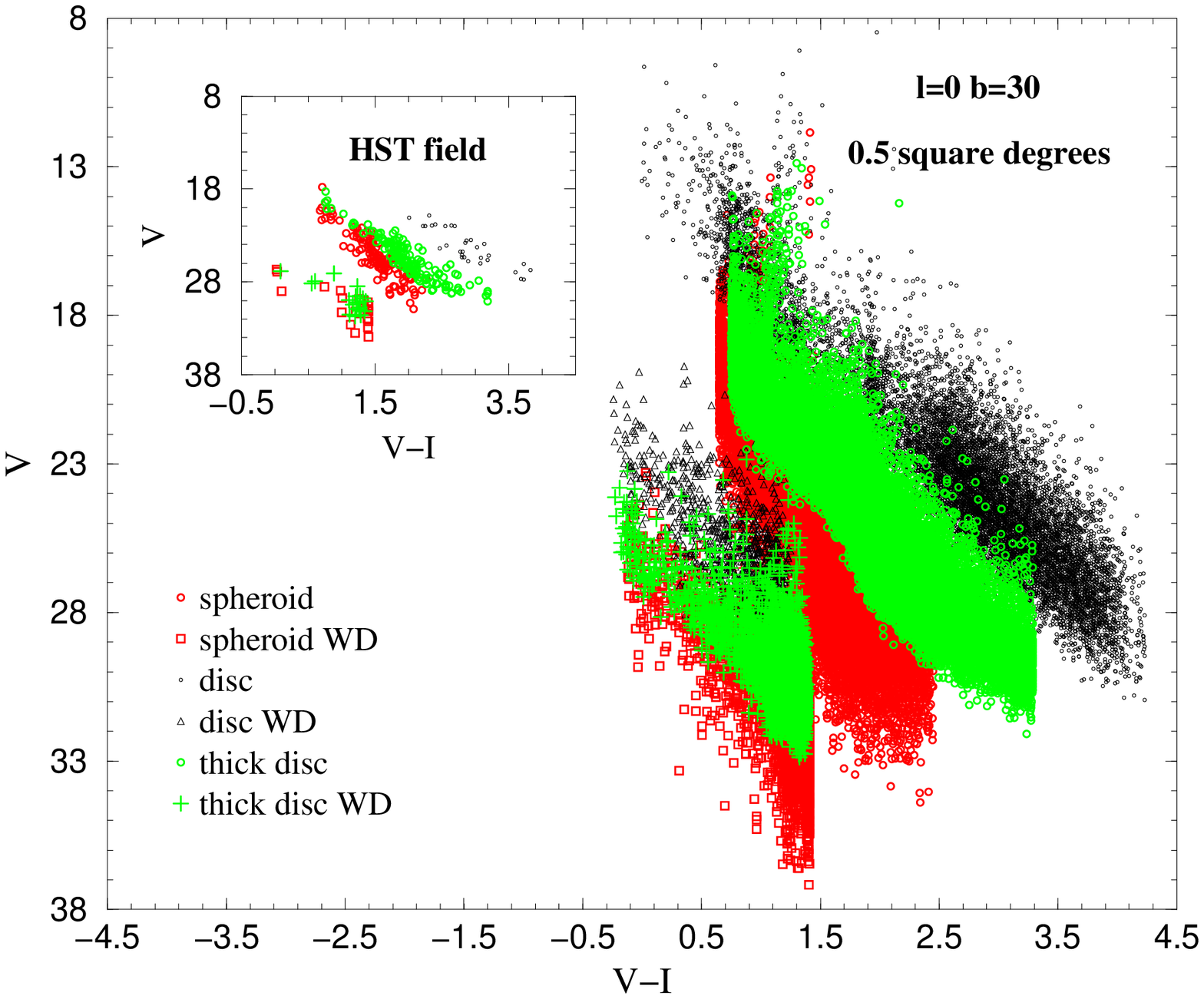}} 
\caption{Theoretical $(V-I, V)$ CMDs for field stars in the three
Galaxy directions: NGP; $l={0}^{\circ}$, $b={50}^{\circ}$;
$l={0}^{\circ}$, $b={30}^{\circ}$; the area is 0.5 square degrees.
Different symbols refer to stars of the various Galaxy populations as
labeled; white dwarfs are separately shown. Note that in our model we
do not introduce artificial colour dispersion simulating observational
spread in colours.}
\end{figure}
%================================================
%==============================================  FIGURA 9
\begin{figure}
\label{HRvr}
\centerline{\epsfxsize= 8 cm \epsfbox{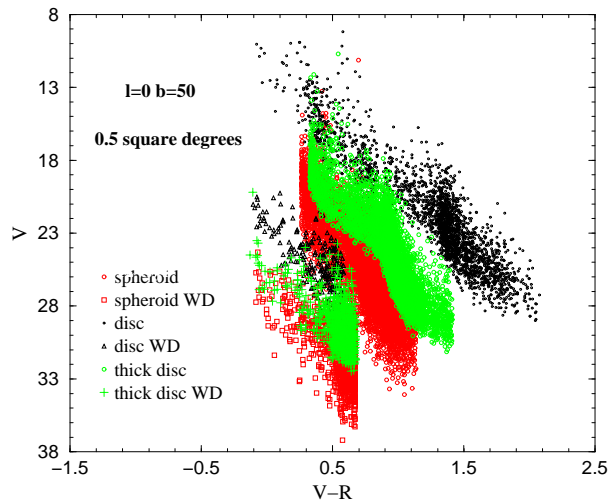}}
\caption{Theoretical $(V-R, V)$ CMD for stars in a field centered at
the Galactic coordinates $l={0}^{\circ}$, $b={50}^{\circ}$  (area of 
0.5 square degrees).  Different symbols refer to stars of the various
Galactic components, as labeled; white dwarfs are separately shown.}
\end{figure}
%================================================
%\caption{Theoretical $(V-I, V)$ CMD for stars in a field of area
% 0.5 square degrees at the Galactic coordinates $l={0}^{\circ}$, $b={30}^{\circ}$.
% Different symbols .......  refers to stars of the various
% Galaxy populations; the white dwarfs are separately shown.}
As discussed above, our model is able to predict various colour
indices for star counts. As an example, Fig. 9 shows the $(V-R, V)$
CMD for the same field of the middle panel of Fig. 8.

The small panels in Fig. 8 show the CMDs obtained for a field of about
6.6 arcmin$^2$ of extension, that is the area generally covered by
{\em{Hubble Space Telescope}} ({\em{HST}}) observations (see e.g. King
et al. 1998).  As expected, at the NGP only a negligible number of
stars is present in such a small field.  Star counts increases moving
from higher to lower latitudes, nevertheless disc population is still
absent at intermediate latitude ($b={50}^{\circ}$), beginning to
appear only at lower latitude ($b={30}^{\circ}$), where there's not
yet evidence of disc WDs.

As well known, at high Galactic latitude the various Galactic components
are expected to contribute to the total CMD with different
colours. While the bluer colours are dominated by spheroid stars and
the redder object belongs to the disc population, the intermediate
colours are dominated by thick disc stars.  As a consequence, the
whole sample of stars in the CMD appears rather uniformly spreaded,
without the ``double peaked'' feature usually predicted by the
two-components models, due to the separation in colour between the disc
and the spheroid population.

We are now in the position to explore the role played by the Galactic
WDs on the star counts.  Figure 10 shows the predicted contribution of
each Galaxy component to the $V$-magnitude distribution of the sample
at the intermediate latitude $b={50}^{\circ}$.
%==============================================  FIGURA 10
\begin{figure}
\label{V050}
\centerline{\epsfxsize= 8 cm \epsfbox{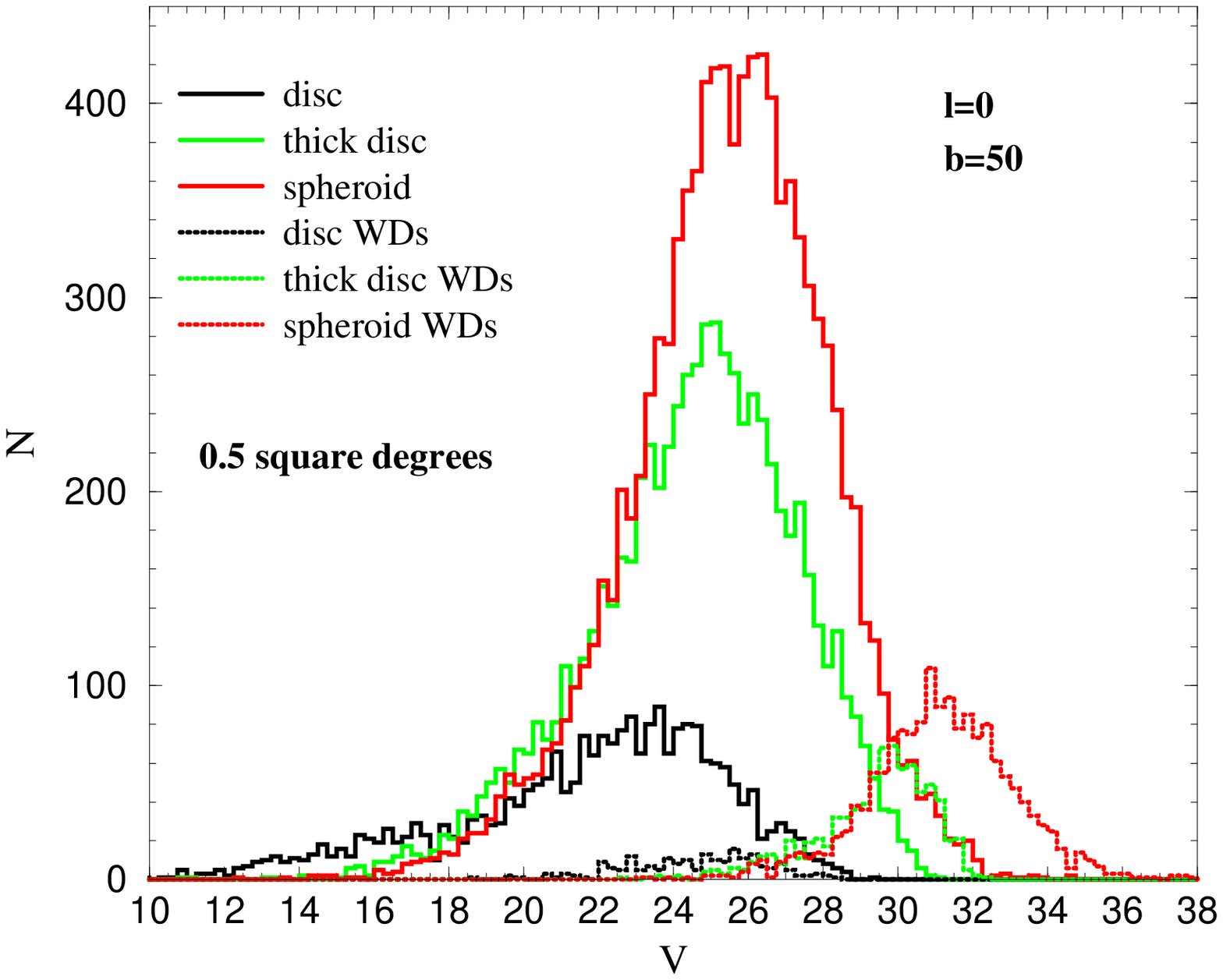}}
\centerline{\epsfxsize= 8 cm \epsfbox{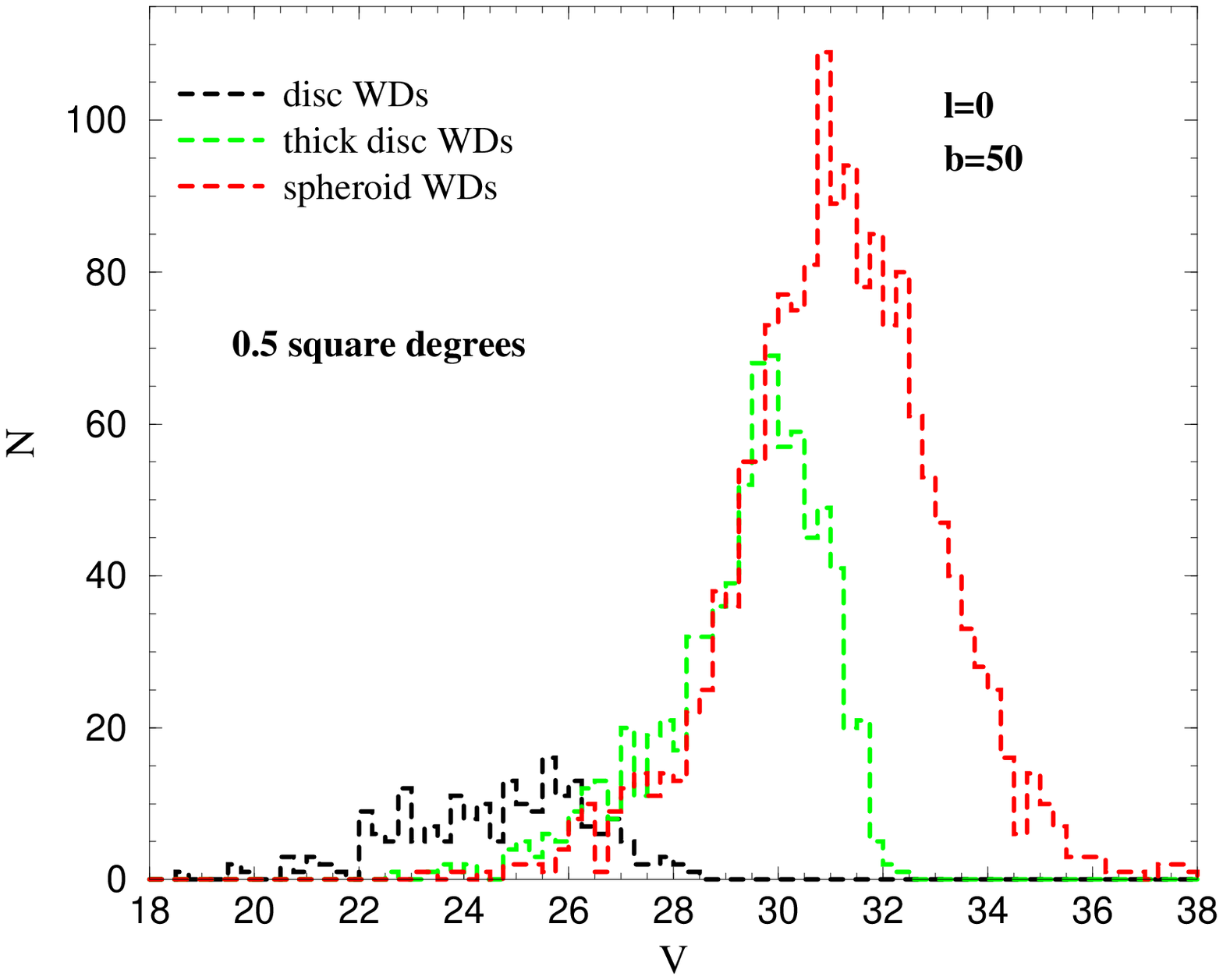}}
\caption{Theoretical apparent-$V$ magnitude distribution of stars in
the field of Fig. 9 ($l={0}^{\circ}$, $b={50}^{\circ}$). Symbols for
the various Galactic populations are labeled. The bottom panel is an
enlargement of the upper one. The adopted spheroid age is 12 Gyr.}
\end{figure}
%================================================
An important finding of our prediction is that observations down to
$V=28$ include almost the whole disc population and disc WDs.
However, only a few percent of thick disc and spheroid WDs are
observable to this luminosities; in fact Fig. 10 shows that the thick
disc WD distribution appears centered at $V\sim 30$, while the
spheroid distribution is centered at $V\sim 31$, with a tail reaching
faint luminosities down to $V\sim 36$.

Figure 11 (upper panel) shows the predicted ($V-I$)-colour distributions
of the Galaxy components in the apparent magnitude range $19<V<27$.
The bottom panel is just an enlargement of the figure displaying the
behaviour of the WD populations.
%==============================================  FIGURA 11
\begin{figure}
\label{VI050}
\centerline{\epsfxsize= 8.5 cm \epsfbox{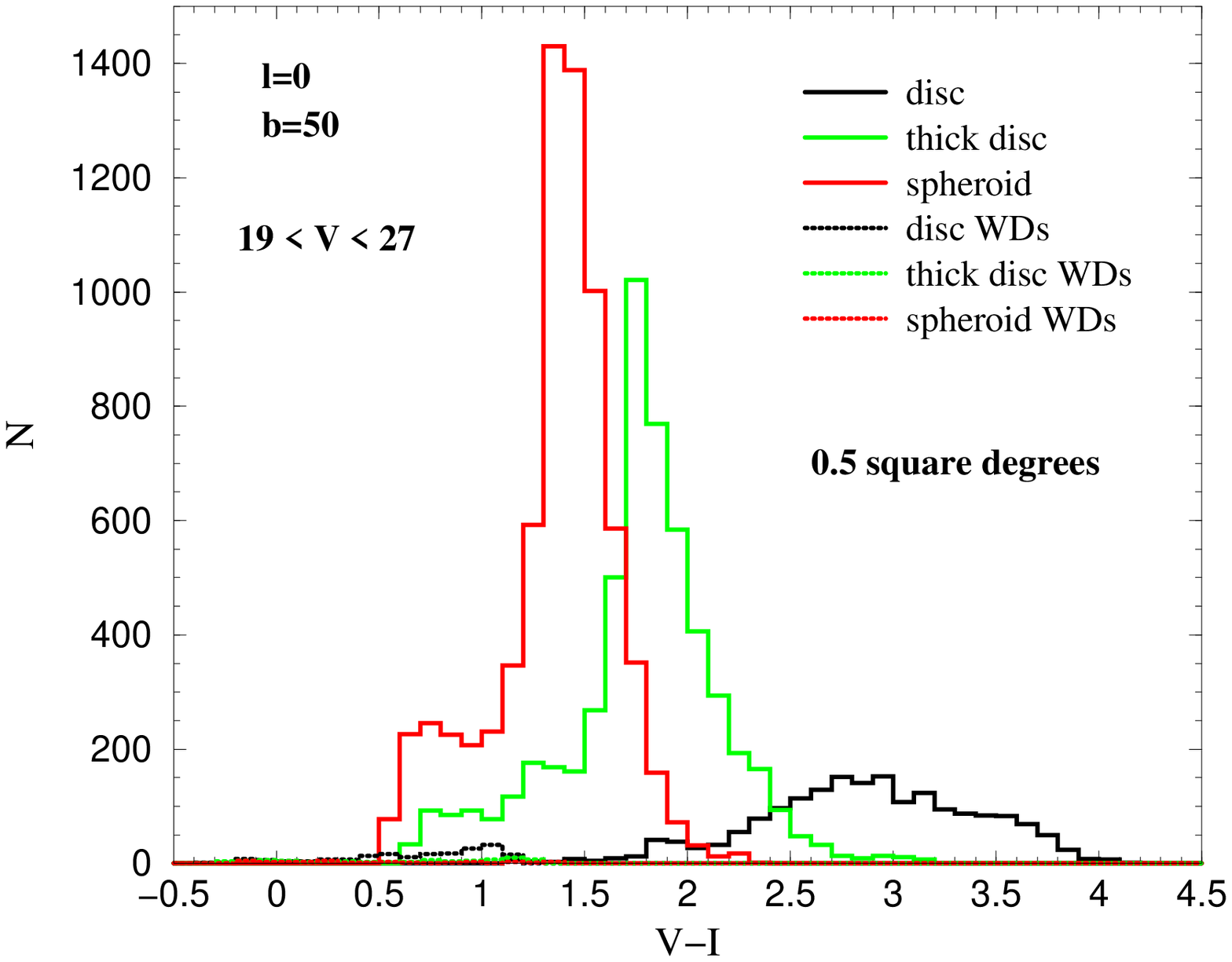}}
\centerline{\epsfxsize= 8 cm \epsfbox{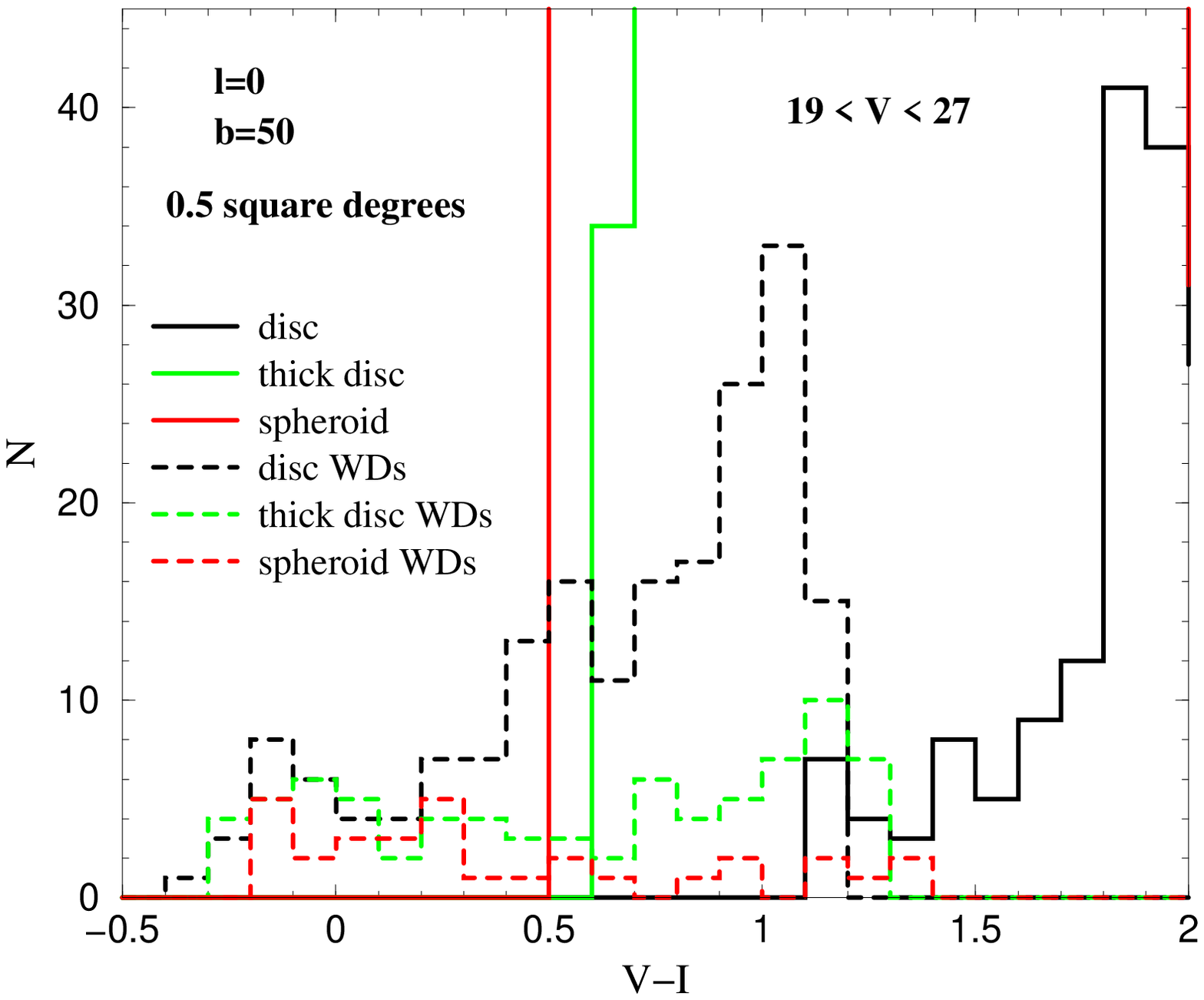}}
\caption{Theoretical ($V-I$)-colour distribution of stars in the field
of Fig. 9 ($l={0}^{\circ}$, $b={50}^{\circ}$). Symbols for the various
Galactic populations are labeled. The bottom panel is an enlargement of
the upper one. Note that in our model we do not introduce artificial
colour dispersion simulating observational spread in colours.}
%Theoretical ($V-I$)-colour distribution of stars in a field of area
% 0.5 square degrees at the Galactic coordinates $l={0}^{\circ}$, $b={50}^{\circ}$.
% Symbols for the various Galaxy populations are labelled in the figure.
\end{figure}
%================================================
The whole WD population takes place at colours bluer then $V-I \sim
1.5$.  Inspection of this figure reveals that there are regions of the
CMDs in which the contribution of white dwarfs to the star counts
seems to be distinguishable from other Galaxy stars. In particular, at
sufficiently blue colours, i.e. $V-I \la 0.5$, and not too high
luminosity ($V \ga 16$), the sample should be constituted exclusively
by WDs.

\section{Conclusions}

We presented a Galactic model able to reproduce star counts and
synthetic colour-magnitude diagrams of field stars from the main
sequence to the white dwarf evolutionary phase for various photometric
bands and Galactic coordinates. The main goal is the introduction of
disc/thick disc and spheroid WD population in an evolutionary
consistent way. To this aim we relied on suitable WD mass-progenitor
mass relations, theoretical WD models and colour transformations. As a
result, we find that the predicted local WD luminosity function
appears in {\em good} agreement with recent observational estimates.
Predictions for CMDs and star counts for the various components in
different fields of the Galaxy are shown. Luckily enough, we find that
the WD population is barely sensitive to a change of theoretical WD
models or to a variation of the adopted relation between the WD mass
and the progenitor mass while, as expected, the results are
significantly affected by the variation of the initial mass function
IMF for masses which could evolve into WDs in a time shorter than the
estimated age of the Universe.

\section*{Acknowledgments}

We warmly thank M. Salaris for making available to us the Salaris et
al. (2000) WD evolutionary tracks. The partial support by MURST within
the ``Stellar Observables of Cosmological Relevance'' project is
acknowledged by the authors.

\label{lastpage}

\end{document}